\documentclass{revtex4}

\usepackage{graphicx}

\usepackage{amssymb}

\def\lsim{\buildrel {\textstyle <}\over {_\sim}}

\begin{document}


\title{Spin-chirality decoupling in Heisenberg spin glasses and related systems}\author{Hikaru Kawamura}
\address{Faculty of Science, Osaka University, Toyonaka 560-0043, Japan}

\begin{abstract}
Recent studies on the spin and the chirality orderings of the three-dimensional  Heisenberg spin glass and related systems are reviewed with particular emphasis on the possible spin-chirality decoupling phenomena.  Chirality scenario of real spin-glass transition and its experimental consequence on the ordering of Heisenberg-like spin glasses are discussed. 
\end{abstract}

\maketitle


\section{Introduction}

Ordering of spin-glass (SG) have been studied quite extensively as a typical example of ``complex'' systems. Experimentally, convincing evidence has now been obtained for the existence of an equilibrium  phase transition at a finite temperature in typical SG magnets, {\it e.g.\/}, canonical SG. The true nature of the SG transition and of the SG ordered state, however, still remains to be at issue \cite{review}.

In theoretical or numerical studies of SG, a simplified model called the Edwards-Anderson (EA) model has widely been used \cite{review}. For the case of the Ising EA model in three dimensions (3D) corresponding to an infinitely strong magnetic anisotropy, it is now well established that the model exhibits an equilibrium SG transition at a finite temperature \cite{review}. One should bear in mind, however, that the magnetic interactions in  many real SG materials are nearly isotropic, being well described by an isotropic Heisenberg model. Although earlier numerical studies on the 3D Heisenberg EA model suggested that the Heisenberg SG exhibited a SG transition only at $T=0$ \cite{Banavar,McMillan,Olive,Matsubara91,Yoshino}, recent numerical studies tend to suggest in common that the Heisenberg SG in 3D exhibits a finite-temperature transition \cite{Kawamura92,Kawamura96,Kawamura98,HukuKawa00,KawaIma,Matsumoto,ImaKawa,HukuKawa05,Matsubara00,Endoh01,Matsubara01,Nakamura,LeeYoung,BY,Picco,Campos}. Yet, the nature of the transition still remains controversial. Obviously, in order to understand the true nature of the experimental SG ordering, it is crucially important to elucidate the nature of the ordering of the 3D Heisenberg SG.

Some time ago, the present author proposed a scenario, a chirality
scenario, for the ordering of real Heisenberg-like SG
\cite{Kawamura92,Kawamura96}. Chirality is a multispin variable
representing the handedness of the noncollinear or noncoplanar
structures induced by spin frustration. A key notion in this scenario is
the ``spin-chirality decoupling'', which might possibly occur in certain
frustrated magnets including the Heisenberg SG. In this article, I wish to review the present status of research on the spin and the chirality orderings of the 3D Heisenberg SG and related systems.

\section{Chirality}

Two types of chirality have been discussed in the literature, a vector chirality and a scalar chirality. The two-component {\it XY\/} spin system ordered in a noncollinear manner possesses a twofold Z$_2$ chiral degeneracy, according as the noncollinear spin structure is either right- or left-handed, in addition to the SO(2) spin-rotation degeneracy. The vector chirality $\kappa$ is defined as a vector product of the two neighboring spins by $\kappa = \sum  S_i\times  S_j$. The sign of its $z$-component tells which chiral state the system takes.

The three-component Heisenberg spin system ordered in a noncoplanar manner also possesses a twofold Z$_2$ chiral degeneracy, 
in addition to the SO(3) spin-rotation degeneracy. The scalar chirality $\chi$ is defined by the product of three neighboring spins by $\chi = S_i\cdot  S_j\times  S_k$.

As is evident from the definition of the local chirality, the chirality
is a composite operator of the spins locally, not independent of the spin. The spin-chirality decoupling, if any, means that, on sufficient long length and time scales, say, beyond a certain crossover length and time scale, chiral correlations might outgrow spin correlations, {\it i.e.\/}, the chirality correlation length gets much longer than the spin correlation length, $\xi$ (chirality) $>>$ $\xi$(spin). 

 In terms of the phase transition, the spin-chirality decoupling might
 lead to either of the following two situations: In one, the spin and
 the chirality might order at the same temperature, say, at zero
 temperature, where there appear two distinct diverging length scales,
 each associated with the spin and with the chirality. More precisely,
 the chirality correlation-length exponent is greater than the spin
 correlation-length exponent, $\nu$ (chirality) $>$ $\nu$ (spin). In the
 other, the spin and the chirality might order at two distinct
 temperatures. With decreasing the temperature, the chirality orders
 first at a higher temperature followed by the spin order at a lower temperature, $T_c$ (chirality) $>$ $T_c$(spin).

\section{The spin-chirality decoupling in regularly frustrated {\it XY\/} antiferromagnets}

We review briefly the spin and the chirality orderings of regularly frustrated {\it XY\/} antiferromagnets. First example is the classical {\it XY\/} (plane rotator) model on the one-dimensional (1D) triangular-ladder lattice. The model is exactly solvable \cite{Horiguchi}. While both the spin and the chirality order only at $T=0$, the associated correlation-length exponents are mutually different.  Indeed, the spin correlation-length exponent is equal to unity $\nu_s = 1$, while the chiral correlation-length exponent is equal to $\nu_\kappa = \infty$, meaning that the chiral correlation length diverges exponentially toward $T=0$ \cite{Horiguchi}. Hence, in this particular 1D model, the spin-chirality decoupling is rigorously shown to occur.

Another example might be the classical {\it XY\/} antiferromagnet on the 2D triangular lattice. Although there had been  some controversy concerning how the chiral Z$_2$ and the spin-rotation SO(2) order in this system, consensus now appears that separate spin and chirality transitions occur successively \cite{MiyashitaShiba,Xu,Capriotti,Lee,Loison,Ozeki}. With decreasing the temperature, the chirality orders first at a higher temperature into the long-range ordered state, while the spin orders at a lower temperature into the quasi-long-range ordered state.

\section{Spin and chirality orderings of the three-dimensional Heisenberg spin glass}

According to the chirality scenario \cite{Kawamura92,Kawamura96}, the 3D Heisenberg SG exhibits the spin-chirality decoupling. With decreasing the temperature, chiral correlations outgrow spin correlations at some crossover temperature $T=T^*$, and at a lower temperature $T=T_{{\rm CG}}$ the chirality exhibits a glass transition into the {\it chiral-glass\/} ordered state without accompanying the standard SG order. The SG transition temperature is lower than the chiral-glass transition temperature, $T_{{\rm SG}}<T_{{\rm CG}}$, $T_{{\rm SG}}$ being either zero or nonzero. The basic picture is summarized in Fig.1 in terms of the temperature dependence of the spin and the chirality correlation lengths  (correlation times).

  \begin{figure}[t]
\includegraphics[scale =0.6]{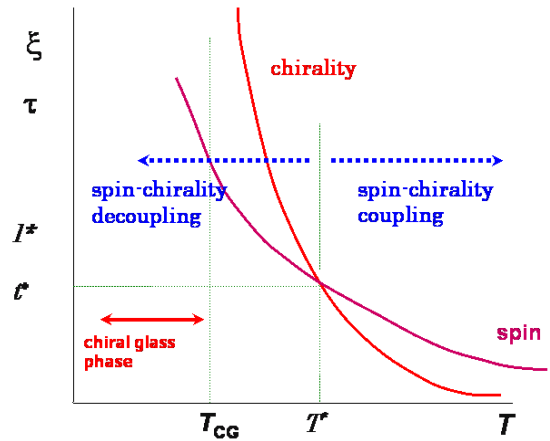}
\caption{
The temperature dependence of the correlation length $\xi$ and the correlation time $\tau$ of the 3D Heisenberg SG both for the spin and for the chirality, expected from the chirality scenario.
}
  	\label{fig-1}
  \end{figure}

As mentioned, the issue of whether the spin-chirality decoupling really occurs in the 3D Heisenberg SG remains controversial. While several numerical results in favor of the occurrence of the spin-chirality decoupling were reported in Refs.\cite{Kawamura98,HukuKawa00,KawaIma,Matsumoto,ImaKawa,HukuKawa05}, a simultaneous spin and chirality transition without the spin-chirality decoupling was claimed in other works \cite{Matsubara00,Endoh01,Matsubara01,Nakamura,LeeYoung,BY,Picco,Campos}.

Here, we wish to report on our recent Monte Carlo results on the spin and the chirality orderings of the 3D Heisenberg SG with the nearest-neighbor $\pm J$ coupling (done in collaboration with Dr. K. Hukushima). Details of the simulation, including the precise definitions of various physical quantities, have been given in Ref.\cite{HukuKawa05}.

In Fig.2, we show the temperature dependence of the spin and the chirality autocorrelation times on a semi-log plot. At higher temperatures spin correlations dominate over chiral correlations, where the system is in the spin-chirality coupling regime. With decreasing the temperature beyond a crossover temperature $T^*$, chiral correlations exceed spin correlations, and the system gets into the spin-chirality decoupling regime at $T\lsim T^*$. Thus, the spin-chirality decoupling appears to be realized in this system.

  \begin{figure}[t]
\includegraphics[scale =0.65]{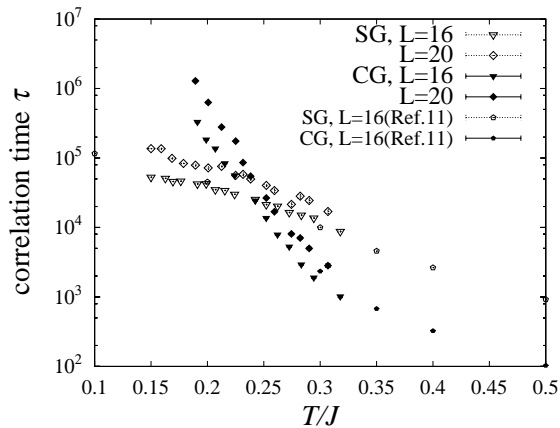}
\caption{
The temperature dependence of the chirality and the spin autocorrelation times   of the 3D $\pm J$ Heisenberg SG on a semi-log plot (taken from Ref.\cite{HukuKawa05}).
}
  	\label{fig-2}
  \end{figure}

In order to clarify the transition behavior of the model, we investigate the Binder ratio of the spin and of the chirality. As shown in Fig.3(a), the chirality Binder ratio exhibits a negative dip which deepens with increasing the system size $L$. The data of different $L$ cross on the {\it negative\/} side of $g_{{\rm CG}}$. These features strongly suggest the existence of a finite-temperature transition in the chiral sector. By extrapolating the dip temperature to $L=\infty$, we get an estimate of the chiral-glass transition temperature $T_{{\rm CG}}\simeq 0.19$. We note that the observed shape of the chirality Binder ratio resembles the one observed in systems exhibiting a one-step replica-symmetry breaking (RSB). The corresponding spin Binder ratio, by contrast, exhibits no signature of a phase transition, no crossing nor merging. Even at $T=T_{{\rm CG}}$, $g_{{\rm SG}}$ stays completely off-critical. Hence, the Binder ratio suggests the occurrence of a chiral-glass transition  at a finite temperature, $T=T_{{\rm CG}}\simeq 0.19$ without accompanying the standard SG order. 

  \begin{figure}[t]
\includegraphics[scale =0.6]{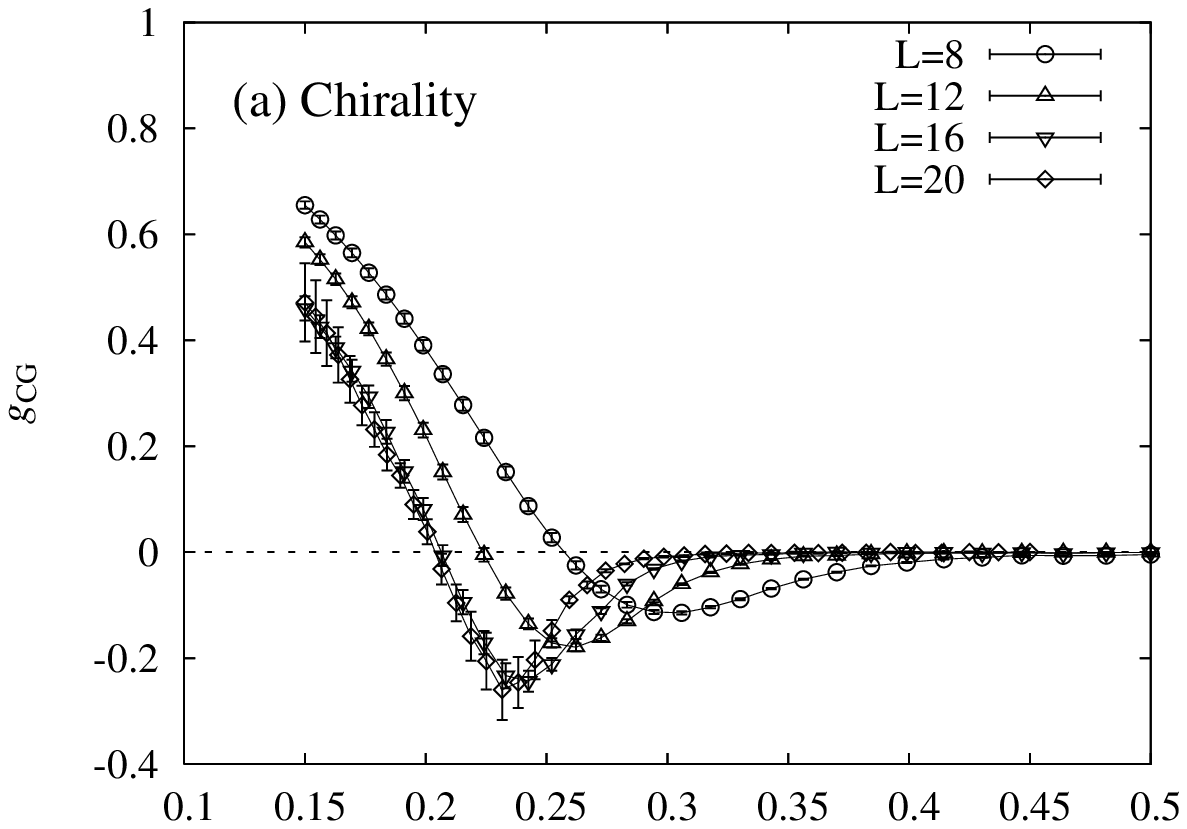}
\includegraphics[scale =0.6]{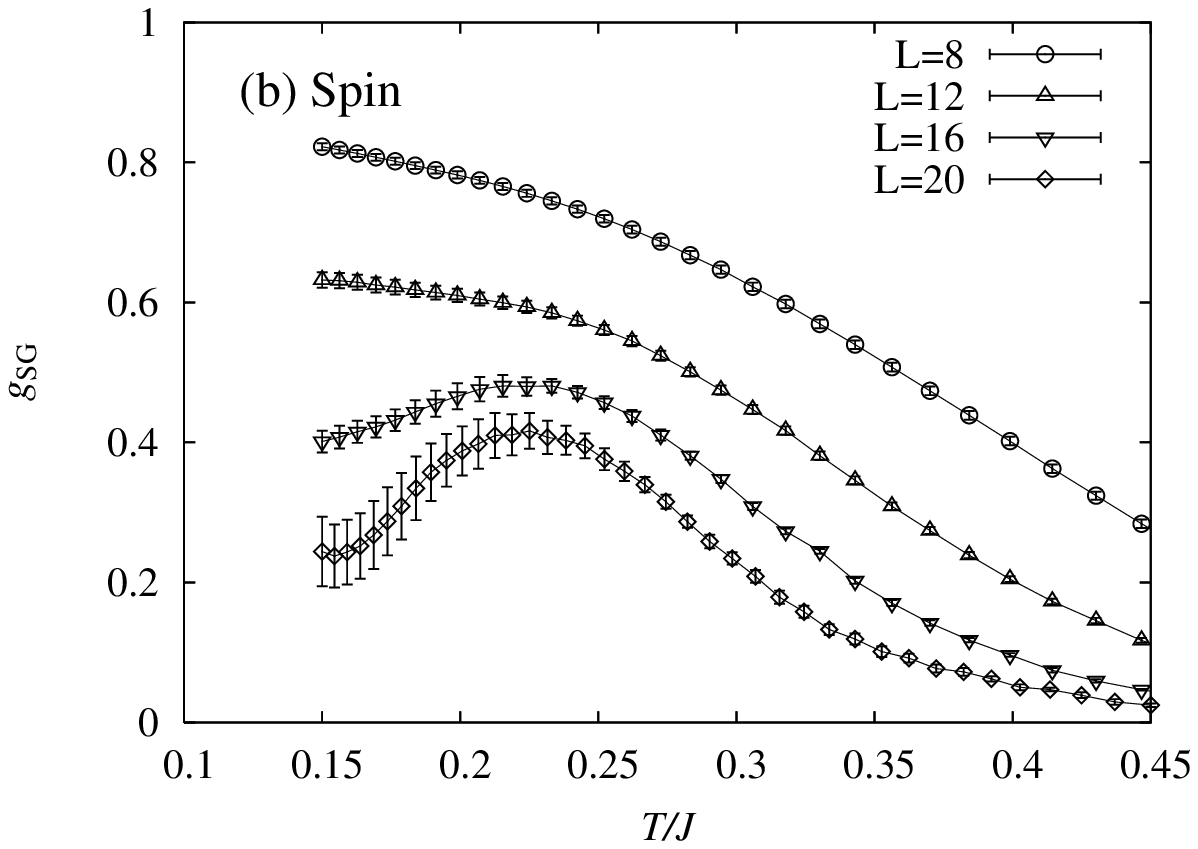}
\caption{
The temperature dependence of the chirality Binder ratio (a), and of the spin Binder ratio (b), of the 3D $\pm J$ Heisenberg SG 
(taken from Ref.\cite{HukuKawa05}).
}
  	\label{fig-3}
  \end{figure}

The overlap distribution function in the ordered state is shown in Fig.4
for both cases of the chirality (Fig.4(a)) and of the spin (Fig.4(b)) at
a temperature $T=0.15$. The chirality overlap distribution $P(q_\chi)$
exhibits symmetric side peaks at $q_\chi =\pm q_\chi^{{\rm EA}}$
corresponding to the long-range chiral-glass order, which grow with
increasing $L$. On top of it,  $P(q_\chi)$ also exhibits a {\it centra peak\/} at $q_\chi =0$, which also grows with increasing $L$. The existence of such a pronounced central peak is a characteristic feature of the system exhibiting a one-step-like RSB, never seen in the Ising SG. The data strongly suggest that the chiral-glass ordered state exhibits a one-step-like RSB. By contrast, the spin overlap distribution $P(q_{{\rm diag}})$, calculated for the diagonal component of the spin-overlap tensor $q_{{\rm diag}}=\sum_\mu q_{\mu \mu}$ ($\mu=x,y,z$), shows an entirely different behavior: Although $P(q_{{\rm diag}})$ exhibits symmetric peaks at finite values of $q_{{\rm diag}}$ for smaller lattices, suggesting the appearance of the SG long-range order, these peaks gradually go away for larger lattices,  and $P(q_{{\rm diag}})$ tends to a single-peak function around  $q_{{\rm diag}}=0$, which is a characteristic of the disordered phase. 

  \begin{figure}[t]
\includegraphics[scale =0.55]{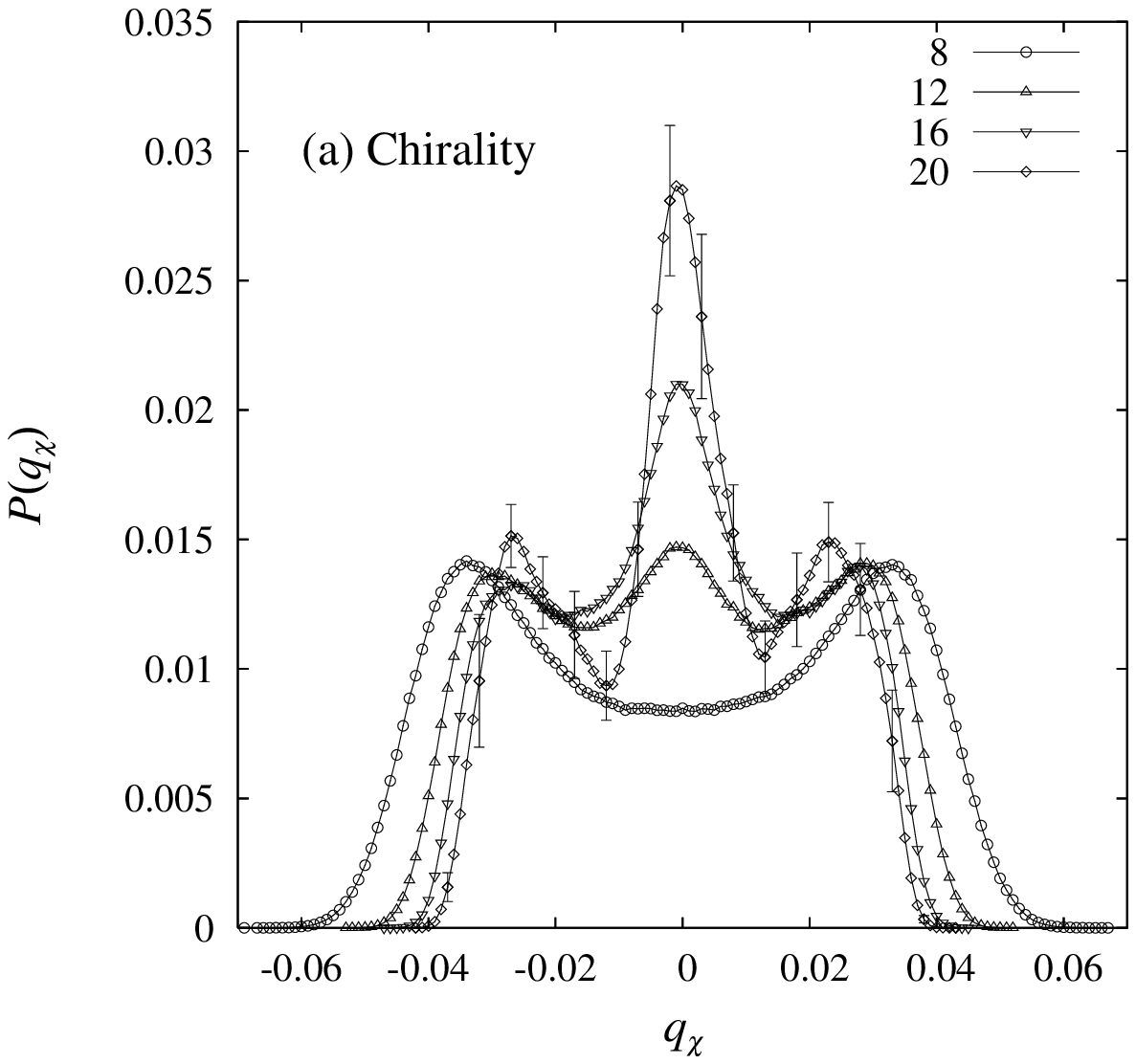}
\includegraphics[scale =0.55]{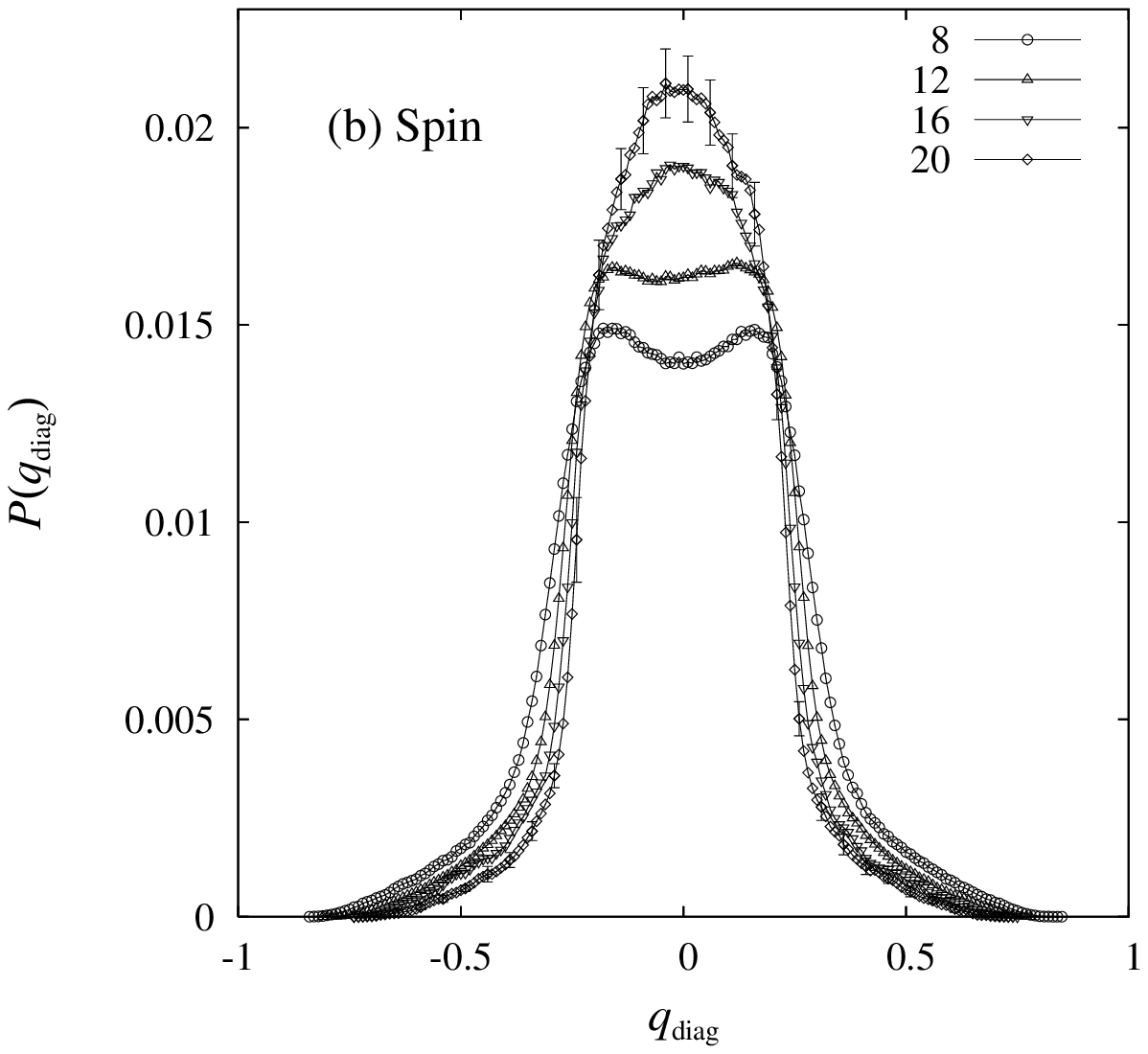}
\caption{
The overlap distribution function for the chirality (a), and for the
   diagonal component of the spin overlap tensor (b), of the 3D  $\pm J$
   ¡¡ Heisenberg SG (taken from Ref.\cite{HukuKawa05}).  The temperature is $T=0.15$ well below the chiral-glass transition point $T\simeq 0.19$.
}
  	\label{fig-4}
  \end{figure}

In Fig.5, we show the dimensionless correlation lengths both for the
chirality and for the spin, $\xi_{{\rm CG}}/L$ and $\xi_{{\rm SG}}/L$,
the data for smaller lattices ($L=8$ and 12) in upper panel and  those for larger lattices ($L=16$ and 20) in lower panel. For smaller lattices, both the spin and the chirality correlation lengths cross at a more or less common temperature, which seems consistent with the observation of Ref.\cite{LeeYoung}. By contrast, for larger lattices, while the chiral correlation length still exhibits a crossing at the expected chiral-glass transition point $T\simeq 0.19$, the spin correlation length does not quite cross any longer, only a merging-like behavior being observed  below $T_{{\rm CG}}$.  Thus, our observation for the spin correlation length is that the crossing tendency is more and more weakened if one goes to larger lattices. It is not clear at the present stage what is a true asymptotic behavior of the spin correlation length for large enough lattices. Since the spin-chirality decoupling, if any, should manifest itself beyond a certain crossover length $L^*$, the behavior of $\xi_{{\rm SG}}/L$ observed here seems consistent with the spin-chirality decoupling with $L^*\simeq 20$. We also note that, although it is sometimes argued that the normalized correlation length would be the best quantity in probing the ordering behavior \cite{LeeYoung}, there exists an occasion where $\xi_{{\rm SG}}/L$ overestimates the ordering tendency \cite{Matsuda}. It is thus very important to examine the ordering behavior of the system by comparing the dimensionless correlation length with various other independent quantities, {\it e.g.\/}, the order parameter, the Binder ratio or the overlap distribution. 

\begin{figure}[t]
\includegraphics[scale =0.6]{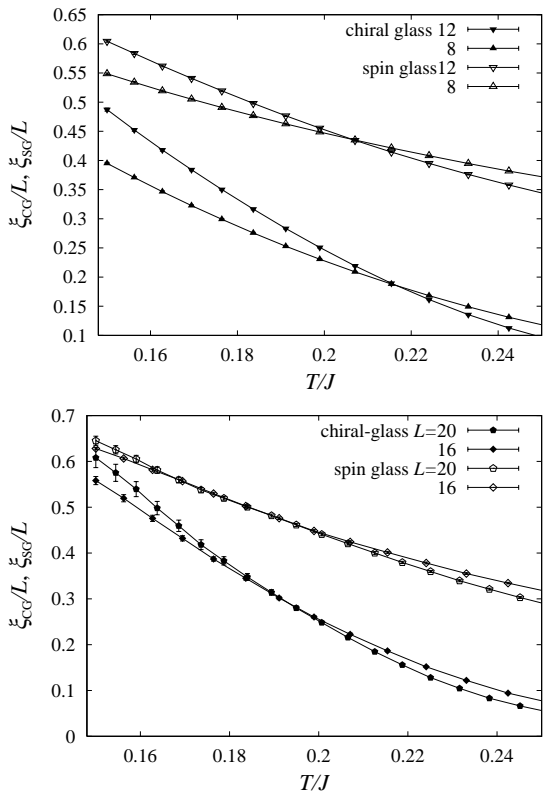}
\caption{
The temperature dependence of the dimensionless spin and chirality correlation lengths of the 3D  $\pm J$ Heisenberg SG, for smaller lattices $L=8$ and 12 (above), and for larger lattices $L=16$ and 20 (below) (taken from Ref.\cite{HukuKawa05}).
}
  	\label{fig-5}
  \end{figure}

Very recently, Campos {\it et al\/} studied the 3D Heisenberg SG with
the Gaussian coupling for larger lattices up to $L=32$, though the
temperature range was limited to just below the transition temperature
\cite{Campos}.  Their data of $\xi_{{\rm CG}}/L$ exhibits a weak
crossing at around $T\simeq 0.15$, whereas those of $\xi_{{\rm SG}}/L$
for larger lattices $L\geq 16$ do not cross in the
investigated temperature range. Hence, at least the raw data of the
dimensionless correlation lengths are consistent with the spin-chirality
decoupling picture. Nevertheless, the authors of Ref.\cite{Campos}
interpreted the data as suggesting a simultaneous spin and chirality
transition of Kosterlitz-Thouless (KT)) type, by invoking a large
correction-to-scaling term. This interpretation, however, seems not persuading. We note that, in the type of systems whose ordered state
exhibits a one-step-like RSB, the corresponding dimensionless
correlation length $\xi(L)/L$  might tend to a finite value even below
 $T_c$ at $L\rightarrow \infty$, {\it not
diverging to infinity\/}, disguising a KT transition, in sharp contrast to the standard system where
$\xi(L)/L\rightarrow \infty$ in the ordered state. This is because, in a one-step RS broken state, the ordered state is quite exotic consisting of many pure states which are mutually dissimilar with vanishing overlaps. 

We also estmate the chiral-glass exponents via the standard finite-size scaling analysis. 
The exponents obtained are $\nu _{{\rm CG}}\simeq 1.2$ and $\eta _{{\rm
CG}}\simeq 0.8$, {\it etc.\/}, which deffer significantly from the
standard 3D Ising SG values, $\nu \simeq 2\sim 3$ and $\eta \simeq -0.35\sim -0.4$ \cite{review}. The results indicate that the chiral-glass transition belongs to a universality class different from the one of the 3D Ising SG. Possible long-range and/or many-body nature of the chirality-chirality interaction might be the cause of this difference.

\section{Chirality scenario for the weakly anisotropic Heisenberg SG}

On assuming that the spin-chirality decoupling occurs in the 3D
isotropic Heisenberg SG, we now ask: What does this mean for real
Heisenberg-like SG where the weak random magnetic anisotropy inevitably
exists ? The chirality scenario claims that the weak random anisotropy
inherent to real SG magnets ``recouples'' the spin to the chirality, and
the chiral-glass transition of the isotropic system is revealed as the
standard SG transition in real weakly anisotropic Heisenberg SG \cite{Kawamura92,Kawamura96}. 

Such a ``spin-chirality recoupling'' can be understood based on a simple symmetry consideration. The isotropic Heisenberg SG possesses both the chiral Z$_2$ symmetry and the spin-rotation SO(3) symmetry, {\it i.e.\/}, Z$_2 \times$ SO(3). Due to the spin-chirality decoupling, only the chiral Z$_2$ is spontaneously broken in the isotropic system at the chiral-glass transition with keeping the SO(3) symmetry unbroken, which leavs the spin to be paramagnetic even below $T_{{\rm CG}}$. Suppose that the weak random anisotropy is added to the isotropic system. It energetically breaks the SO(3) symmetry with keeping the chiral Z$_2$ symmetry. (Note that the invariance under the spin inversion $S \rightarrow - S$, which flips the chirality, is kept in the presence of the random magnetic anisotropy.)  Since the chiral Z$_2$ has already been decoupled from the SO(3) in the isotropic system, it would be natural to expect that the Z$_2$ chiral-glass transition of the anisotropic system occurs essentially in the same manner as that of the isotropic system. As soon as the Z$_2$ chiral-glass transition takes place, however, there is no longer any symmetry left in the anisotropic system, which forces the spin to order below $T_{{\rm CG}}$. This is a spin-chirality recoupling due to the magnetic anisotropy. 

The situation might be summarized in the schematic phase diagram in the anisotropy ($D$) versus temperature ($T$) plane of Fig.6. In the isotropic limit $D=0$, due to the spin-chirality decoupling there, the chiral-glass transition occurs at a temperature higher than the SG transition temperature, $T_{{\rm CG}} > T_{{\rm SG}}$. A crucial observation is that the SG (simultaneously chiral-glass) transition of the anisotropic system with $D>0$ is a continuation of the chiral-glass fixed point of the isotropic $D=0$ system, {\it not\/} a continuation of the SG fixed point of the isotropic system. The SG transition of real Heisenberg-like SG with weak random anisotropy is governed by the same chiral-glass fixed point all the way along the transition line, including both $D=0$ and $D>0$. In this way, the $D\rightarrow 0$ limit is not singular, and  there is no Heisenberg-to-Ising crossover in the SG critical properties even in the $D\rightarrow 0$ limit.

  \begin{figure}[t]
\includegraphics[scale =0.55]{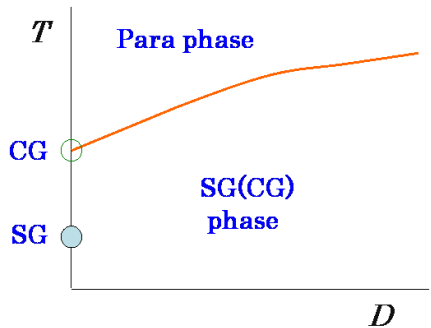}
\caption{
The schematic phase diagram of the weakly anisotropic Heisenberg spin
   glass in the anisotropy ($D$) versus temperature ($T$) plane. CG and SG
   stand for the chiral glass and the spin glass, respectively.
}
  	\label{fig-6}
  \end{figure}

Such a picture leads to the following interesting predictions on the properties of experimental Heisenberg-like SG. i) The SG transition temperature $T=T_g$ depends on the anisotropy $D$ in a regular manner, as $T_g(D)\sim T_{{\rm CG}}(0)+cD + \cdots $ ($c$ is a numerical constant).  ii) The SG critical exponents are given by the chiral-glass critical exponents of the isotropic system, which differ significantly from the 3D Ising SG exponents. They are  $\beta \sim 1$, $\gamma \sim 2$, $\delta \sim 3$ and $\eta $ positive. Furthermore, even the weakly anisotropic SG does not show Heisenberg-to-Ising crossover in its critical behavior. iii) The SG ordered state of Heisenberg-like SG exhibits a one-step-like RSB. As a corollary of this, experimental Heisenberg-like SG is expected to exhibit an equilibrium SG transition even under magnetic fields as an RSB transition. This provides an interesting, and somewhat unexpected possibility that an in-field ordering behavior of the weakly anisotropic Heisenberg-like SG might entirely differ from that of the strongly-anisotropic Ising SG. For the latter, recent theoretical studies suggest that there is no equilibrium in-field SG transition \cite{Katzgraber,Takayama}. iv) The magnetic phase diagram of experimental Heisenberg-like SG resembles the one of the corresponding mean-field model: In the high field regime, the SG transition line behaves as the mean-field Gabay-Thouless (GT) line with an exponent $1/2$, {\it i.e.\/}, $H_g\sim |T_g(H)-T_g(0)|^{1/2}$ \cite{KawaIma}, while, in the low field regime, it behaves as the mean-field de Almeida-Thouless (AT) line, {\it i.e.\/}, $H_g\sim |T_g(H)-T_g(0)|^{(\beta_{{\rm CG}}+\gamma _{{\rm CG}})/2}$, where $\beta_{{\rm CG}}$ and $\gamma _{{\rm CG}}$ are the corresponding chiral-glass exponents of the isotropic system. Since one has $\beta_{{\rm CG}}\sim 1$ and $\gamma _{{\rm CG}}\sim 2$, one gets an exponent close to $3/2$, which happens to be close to the corresponding AT-line exponent $3/2$.

Basically, these predictions from the chirality scenario are compared favorably with the existing experimental data for real Heisenberg-like SG including canonical SG. Namely, i) experimentally observed anisotropy dependence of the SG transition temperature is close to one expected from the chirality scenario \cite{Fert}. ii) The SG critical exponents observed by various researchers for canonical SG, $\beta\simeq 1$, $\gamma\simeq 2$ and $\eta \simeq 0.5$, are in good agreement with each other, but deviate significantly from the Ising SG values: See, {\it e.g.\/}, Ref.\cite{Levy} and references cited therein. By contrast, these experimental values agree very well with the chiral-glass values. Furthermore, the absence of the expected Heisenberg-to-Ising crossover in the critical behavior of the weakly anisotropic Heisenberg-like SG is the property which has puzzled SG researchers for years \cite{review}. iii) Experimentally, the in-field properties are often significantly different between in the Ising-like SG and in the Heisenberg-like SG \cite{Campbell}. 
iv) While the experimentally determined magnetic phase diagram of the
Heisenberg-like SG is often well described by the mean-field phase
diagram including the GT and AT lines, the true origin of this
coincidence has long been a mystery. Remember that, generally
mean-field theory does not give exponent values of real systems correctly. In contrast, the chirality scenario gives a natural alternative explanation for the apparently mean-field-like phase diagram widely observed in experimental Heisenberg-like SG.

The most stringent experimental test of the chirality scenario would be to directly measure the chirality, particularly, the chiral susceptibility $X_\chi$ and the nonlinear chiral susceptibility $X_{\chi}^{nl}$. This has long remained to be an extremely difficult task, since the chirality is a higher-order quantity in spins, cubic in spins. Recently, however, it has been recognized that the chirality might be measurable by using the anomalous Hall effect as a probe. In fact, G. Tatara and the present author analyzed the chirality contribution to the anomalous Hall effect of metallic SG based on the perturbation analysis \cite{TataraKawamura,Kawamura03}. The anomalous Hall coefficient $R_s$ is then given by
\begin{eqnarray}
R_s &=& \rho / M \nonumber\\
    &=& - \left( A\rho + B\rho^2 \right) - CD \left( X_\chi + X_{\chi}^{nl}(DM)^2 + \cdots \right). 
\end{eqnarray}
It consists of two kinds of terms. The first part is the standard
contribution to the anomalous Hall effect, which is proportional to the
resistivity $\rho$ or its squared $\rho^2$. Since the resistivity does
no show any anomaly at $T_g$, this first part can be regarded as a
regular background. The second part is the chirality contribution, which
is proportional to the chiral susceptibility $X_\chi$. It even contains
the information  of the nonlinear chiral susceptibility  $X_{\chi}^{nl}$
as a higher-order contribution. 

Inspired by this theoretical suggestion, several experimental groups
tried to measure the chirality contribution to the anomalous Hall effect
in metallic SG. These measurements observed a sharp cusp-like anomaly at
$T=T_g$ in the temperature dependence of $R_s$
\cite{Sato,Taniguchi04,Campbell04,Taniguchi06}, followed by the
deviation between the field-cooled and the zero-field-cooled data below
$T_g$ \cite{Taniguchi04,Campbell04}. Furthermore, Taniguchi {\it et
al\/} very recently observed a singular behavior of the nonlinear chiral
susceptibility at $T=T_g$ characterized by the exponent $\delta_{{\rm
CG}}\simeq 3$, which is rather close to the corresponding chiral-glass
exponent \cite{Taniguchi06}. All these observations indicate that the chirality  in
metallic SG indeed exhibits  a strong anomaly at the SG transition,
providing strong experimental support to the chirality scenario of SG
transition. We stress that, if the order parameter of the SG transition
were not the  chirality but were the spin itself as in the case of the
mean-field (Sherrington-Kirkpatrick) Heisenberg SG,  the chiral
susceptibilities would not exhibit such a strong singularity: For
example, the nonlinear chiral susceptibility of the Heisenberg SK model
does not diverge at $T_g$ \cite{ImaKawa03}. This is simply due to the fact
that the chirality is a composite operator, being of higher-order in the
spin. Hence, in the absence of the spin-chirality decoupling, a 
power-counting argument should apply as a first-order approximation,
which leads to $\nu_{{\rm CG}}=\nu_{{\rm SG}}$, $\beta_{{\rm
CG}}\simeq 3\beta_{{\rm SG}}$ and $\gamma_{{\rm CG}}\simeq \gamma_{{\rm
SG}}-4\beta_{{\rm SG}}$, {\it etc\/}. If one substitutes here
the experimental SG exponents for canonical SG, $\beta_{{\rm SG}}\simeq
1$ and $\gamma_{{\rm SG}}\simeq 2$, one gets the chiral-glass
susceptibility exponent $\gamma_{{\rm CG}}\simeq
-2 < 0$, meaning that the nonlinear chiral susceptibility should not diverge !

\section{Summary}

Recent studies on the spin and the chirality orderings of the 3D
Heisenberg SG and related systems were reviewed, with particular
emphasis on the possible spin-chirality decoupling phenomena. Our Monte
Carlo results support the view  that the 3D isotropic Heisenberg SG
exhibits a spin-chirality decoupling, {\it i.e.\/}, a finite-temperature
chiral-glass transition not accompanying the standard SG order.
Chirality scenario of real spin-glass transition and its experimental
consequence on the ordering of real Heisenberg-like SG were
discussed. The scenario appears to explain some of the long-standing
experimental puzzles concerning the Heisenberg-like SG, thereby getting
some support from experiments, particularly from the recent Hall
measurements.  Then, the chirality might be a ``missing link'', playing a crucial role in the ordering of SG.

The author is thankful to Dr. K. Hukushima, Dr. G. Tatara, Dr. D. Imagawa and Mr. A. Matsuda for their collaboration, and to Dr. I. Campbell, Dr. T. Taniguchi, Dr. E. Vincent, Dr. M. Ocio, Dr. H. Yoshino, Dr. M. Picco, Dr. M. Sato, and Dr. H. Takayama for useful discussion.

\end{document}